\documentclass[prb,nofootinbib,twocolumn,superscriptaddress]{revtex4} 

\usepackage{graphicx}
\usepackage{dcolumn}
\usepackage{bm}
\usepackage{threeparttable}
\usepackage{times}
\usepackage{mathptmx}
\usepackage{lscape}
\usepackage{natbib}
\usepackage{amsmath}
\usepackage{amssymb}
\usepackage{braket}
\usepackage{comment}
\usepackage{color}

\def\degree{\kern-.2em\r{}\kern-.3em}

\begin{document}

\title{ Thermodynamic Interpretaion of Entanglement in Canonical Nonlinearity  }

\author{Koretaka Yuge}
\affiliation{
Department of Materials Science and Engineering,  Kyoto University, Sakyo, Kyoto 606-8501, Japan\\
}%

\begin{abstract}
{ For classical discrete system under constant composition, typically reffered to as substitutional alloys, canonical average acts as nonlinear map $\phi$ from a set of potential energy surface $\mathbf{U}$ to that of microscopic configuration in thermodynamic equilibrium, $\mathbf{Q}$, which is called ``canonical nonlinearity (CN)''. On statistical manifold, at any given configuration, $\phi$ can be divided into the sum of local and non-local contribution in terms of Kullback-Leibler (KL) divergence, where the former has strong positive correlation with time evolution of the nolinearity (NOL) on configuration space (called ``anharmonicity in structureal degree of freedoms (ASDF)'', while the latter, corresponding to entanglement in SDFs, does exhibit clear correlation with the ASDF. On the other hand, our recent work bridge the different concepts of NOL on configuration space and statistical manifold through stochastic thermodynamics. While the work successfully provides clear relationships between the changes in total NOL through system transition and heat transfer, thermodynamic interpretaion of how the entanglment in SDFs contributes to thermodynamic functions, is totally unclear due mainly to its non-trivial, non-local character. The present study tuckle this problem, deriving upper bound for the entanglement for any given transition in terms of the mutual information, heat transfer and free energy. The present thermodynamic interpretaion will provide quantitative description of how the entanglement in SDFs is dominated by configuration of ground-state structures on configuration space.

}
\end{abstract}


\maketitle

\section{Introduction}
For classical discrete systems under constant composition (typically referred to as substitutional alloys), microscopic configuration under thermodynamic equilibrium can be typically provided by the canocal average under given coordination of $\left\{ q_{1}, \cdots, q_{f} \right\}$:
\begin{eqnarray}
\Braket{q_{k}}_{Z} = Z^{-1} \sum_{i} q_{k}^{\left( i \right)} \exp \left( -\beta U^{\left( i \right)} \right),
\end{eqnarray}
where $Z$ denotes partition function, $f$ the structural degree of freedoms (SDFs), $\beta$ inverse temperature, $U$ potential energy, and summation is taken over all possible configurations.

When we prepare conmplete orthonormal basis function such as generalized Ising model (GIM),\cite{ce} $U$ for any configuration can be exactly written by
\begin{eqnarray}
U^{\left( i \right)} = \sum_{j} \Braket{U|q_{j}} q_{j}^{\left( i \right)},
\end{eqnarray}
where $\Braket{\quad|\quad}$ represents inner product, i.e., trace over configuration space. 
The above equations naturally leads to the concept that canonical average acts as a map $\phi$ from potential energy surface $\vec{U} = \left( \Braket{U|q_{1}},\cdots,\Braket{U|q_{f}} \right) $ to equilibrium configuration $\vec{q}= \left( \Braket{q_{1}}_{Z},\cdots,\Braket{q_{f}}_{Z} \right)$, namely
\begin{eqnarray}
\phi:\quad \vec{U} \mapsto \vec{q}.
\end{eqnarray}
Generally, $\phi$ is a complicated nonlinear map (we call it as ``canonical nonlinearity (CN)''). Therefore, various theoretical approaches have been proposed to approximately predict equilibrium properties including Metropolis alogorism, entropic sampling and Wang-Landau sampling.\cite{mc1,mc2,wl} Although these existing approaches enable efficient and accurate thermodynamic prediction, origin of the CN in terms of the lattice geometry has not been well addressed so far.

Meanwhile, we recently make several significant progress for interpretation of the CN: (i) The CN can be quantified by the special vector field on configuration space called ``anharmonicity in the SDF (ASDF)'' on individual configuration,\cite{asdf} which purely depends on lattice geometry and is independent of temperature and many-body interaction, (ii) The concept of CN as ASDF can be extended to statistical manifold including further non-local nonliear information, deriving that the CN on each configuration can be divided into the sum (in terms of Kullback-Leibler divergence) of three contribution from deviation from Gaussian constraint (dG), entanglement in SDF (which corresponds to nonseparable character in SDF: NS), and nonadditivity in NS.\cite{ig} (iii) We bridge the above concepts of CN on two different space of configuration space and statistical manifold through stochastic thermodynamics introducing special operation along the transition path. 
These studies enables translation from CN evolution to transition for thermodynamic system, clarifying the relationships between changes in CN as KL divergene and heat transfer through thermodynamic transition: i.e., thermodynamic interpretaion of CN is successfully achieved.\cite{st} 
Our recent work further find that the CN from dG contribution exhibit strong positive correlation with ASDF due to reflecting the local CN information, while that from NS, reflecting non-local CN information, does not exhibit effective correlation with ASDF.\cite{cnlgm} 

As described, although our previous works enable multilateral thermodynamic properties of the CN evolution, thermodynamic interpretaion especially of partial contribution from NS is far from understood. The present study tackle this problem, providing effective upper bound for averaged NS through transition in terms of thermodynamic functions with specially introduced operations. The details are shown below.

\section{Concepts and Derivations}
\subsubsection*{Vector field representation of CN}
We here briefly explain the concept of ASDF, the vector field to provide CN information for given configuration $\vec{q}$. ASDF is defined as 
\begin{eqnarray}
\label{eq:asdf}
\vec{A}\left( \vec{q} \right) = \left\{ \phi\left( \beta \right) \circ \left( -\beta\cdot\Gamma \right)^{-1} \right\}\cdot\vec{q} - \vec{q},
\end{eqnarray}
where $\Gamma$ denotes $f\times f$ covariance matrix for configurational density of states (CDOS) before applying many-body interaction to the system. The ASDF purely depends on lattice (and composition), and is independent of temperature and energy: We can know the CN as ASDF without any thermodynamic information, i.e., depending only on configurational geometry.  $A$ takes zero vector when $\phi$ is a linear map 
around $\vec{q}$. Bidirectional stability character between $\vec{U}$ and $\vec{q}$ can be quantitatively given by the logarism of the sum of divergence and Jacobian for $\vec{A}$. Therefore, $\vec{A}$ is a natural measure of CN defined on individual configuration $\vec{q}$. 

\subsubsection*{CN on statistical manifold}
The concepts of ASDF for CN can be naturally extended to the statistical manifold. We have shown that through generalized Pythagorean theorem and dual orthogonal foliation, ASDF-originated CN of $D_{\textrm{KL}}^{\textrm{CN}}$ can be divided into the sum of three contributions, namely,
\begin{eqnarray}
\label{eq:bridge1}
D_{\textrm{CN}} = D_{\textrm{dG}} + D_{\textrm{NS}} + \Delta D_{\textrm{NAD}},
\end{eqnarray}
where $D$ denotes KL divergence, $D_{\textrm{dG}}$ corresponds to contribution from deviation from Gaussian contraint, $D_{\textrm{NS}}$ from NS in SDF, and $\Delta D_{\textrm{NAD}}$ from nonadditivity in NS. $\Delta _{\textrm{NAD}}$ takes non-zero value only for multicomponent $\left( R\ge 3 \right)$ system with special sets of multisite correlations, and we here consider the case of NAD taking zero.

\subsubsection*{Thermodynamics for CN}
We have measured the CN on two different quantity of ASDF and KL divergence repsectively defined on different spaces, which is bridged through stochastic thermodynamics. We here just briefly show the basic concept and major results, and more details are given in our previous paper. 

Equation \eqref{eq:asdf} straightforwardly leads to considering CN as a time evolution of of the discrete dynamical system:
\begin{eqnarray}
\vec{q}_{t+1} = \vec{q}_{t} + \vec{A}\left( \vec{q}_{t} \right).
\end{eqnarray}
Since ASDF takse ``avearage'' information of CN (by taking canonical average in Eq.~\eqref{eq:asdf}), its deterministic evolution can be naturally extended to stochastic evolution by directly considering the caonical distribution itself, leading to the master equation for CN:
\begin{eqnarray}
\label{eq:master}
\frac{d}{dt} P\left( \vec{q}_{k} \right) = \sum_{i} \left[ R\left( \vec{q}_{k}|\vec{q}_{i} \right) P\left( \vec{q}_{i} \right) - R\left( \vec{q}_{i} | \vec{q}_{k} \right) P\left( \vec{q}_{k} \right) \right],
\end{eqnarray}
where $P\left( \vec{q}_{k} \right)$ denotes probability that system takes state $\vec{q}_{k}$ at time $t$ (hereinafter, $P'$ denotes probablity at time $t+1$), and $R\left( \vec{q}_{k}|\vec{q}_{i} \right)$ denotes transition rate from state $\vec{q}_{i}$ to $\vec{q}_{k}$ given by
\begin{eqnarray}
R\left( \vec{q}_{k} | \vec{q}_{i} \right) = Z_{i}^{-1} g\left( \vec{q}_{k} \right) \exp \left[ -\beta \left( \vec{q}_{k}\cdot \vec{V}_{i} \right) \right],
\end{eqnarray}
where $g\left( \vec{q} \right)$ denotes CDOS, and we define partition function for $\vec{q}_{i}$ as $Z_{i}$ given by
\begin{eqnarray}
Z_{i} = \sum_{\vec{q}} g\left( \vec{q} \right) \exp \left[ -\beta \left( \vec{q}\cdot \vec{V}_{i} \right) \right].
\end{eqnarray}
$V_{i}$ is given by 
\begin{eqnarray}
\vec{V}_{i} = \left( -\beta\cdot \Gamma \right)^{-1} \vec{q}_{i},
\end{eqnarray}
so that ensemble average of the stochastic evolution satisfies the original dynamical system, namely,
\begin{eqnarray}
\sum_{\vec{q}_{B}} R\left( \vec{q}_{B}|\vec{q}_{A} \right) \cdot \vec{q}_{B} = \vec{q}_{A} + \vec{A}\left( \vec{q}_{A} \right).
\end{eqnarray}

From above discussions, since evolution of CN satisfies stochastic Markovian process, corresponding thermodynamic (and other related) functions can be  introduced through transition from state $\vec{q}_{A}$ to $\vec{q}_{B}$:
\begin{eqnarray}
\Delta S_{\textrm{sys}} &=& \ln\frac{P\left( \vec{q}_{A} \right) }{P'\left( \vec{q}_{B} \right) } \nonumber \\
\Delta S_{\textrm{b}} &=& \ln\frac{R\left( \vec{q}_{B}|\vec{q}_{A} \right)}{R\left( \vec{q}_{A}|\vec{q}_{B} \right) } \nonumber \\
\Delta d_{0} &=& - \ln\frac{g\left( \vec{q}_{A} \right)}{ g\left( \vec{q}_{B} \right)} \nonumber \\
\Delta F &=& \left( -\beta^{-1}\ln Z_{B} \right) - \left( -\beta^{-1}\ln Z_{A} \right) \nonumber \\
\sigma &=& \Delta S_{\textrm{sys}} + \Delta S_{\textrm{b}},
\end{eqnarray}
where $\Delta S_{\textrm{sys}}$ and $\Delta S_{\textrm{b}}$ respectively denotes system and bath entropy change, $d_{0}$ change in ``stochastic eigen nonlinearity (SEN)'', $F$ free energy and $\sigma$ entropy production. 
To elucidate relationships between CN and thermodynamic function, we measure the above functions from ``Gaussian system'', where its CDOS takes multidimensional Gaussian with covariance matrix $\Gamma$ same as the practical system, providing globally linear map of $\phi$ for any given $\vec{U}$. Hereinafter, we employ tilde $\tilde{M}$ as $M$ measured from that of Gaussian system. 

Under these preparations, we derive the following relationship:
\begin{eqnarray}
\label{eq:unify}
\Delta D_{\textrm{CN}} - \left[ 2\Delta\tilde{d}_{0} - \Braket{\Delta\tilde{d}_{0} }_{AB} \right] = \beta Q,
\end{eqnarray}
where $\Braket{\quad}_{AB}$ is called ``vicinity average (VA)'' corresponding to special average around the transition, and $Q$ denotes heat gain for system. Equation \eqref{eq:unify} clarifies that changes in nonlinearity on statistical manifold (l.h.s.) measured from SEN is identical to system heat gain through  through system state transition from $A$ to $B$ on configuration space. 

\subsubsection*{Thermodynamic Interpretaion of Entanglement in CN}
Although Eq.~\eqref{eq:unify} provides new thermodynamic insight into total CN, how partial contribution, i.e., $D_{\textrm{dG}}$ and $D_{\textrm{NS}}$, indivudually relates to thermodynamic functions, is still totally unclear. Since our recent work on real systems find that 
$D_{\textrm{dG}}$ exhibit (trivial) strong correlation with ASDF while $D_{\textrm{NS}}$ exhibit non-trivial correlation with ASDF, the present study focus on the behavior of $D_{\textrm{NS}}$ in terms of thermodyamics. 

The difficulty to provide thermodynamic interpretaion for $D_{\textrm{NS}}$ is that its underlying CDOS is unclear, or even if clarified, it is not unique (i.e., generally, depending on configuration). This difficulty comes from the fact that $D_{\textrm{NS}}$ corresponds to KL divergence from canonical distribution for practical system and its separable distribution (i.e., product of its marginal distribution), leading to that corresponding $\vec{V}_{i}$ is indefinite, i.e., we cannot straightforwardly employ stochastic evolution of $D_{\textrm{NS}}$ contribution as master equation of 
Eq.~\eqref{eq:master}.

\begin{figure}[h]
\begin{center}
\includegraphics[width=0.88\linewidth]{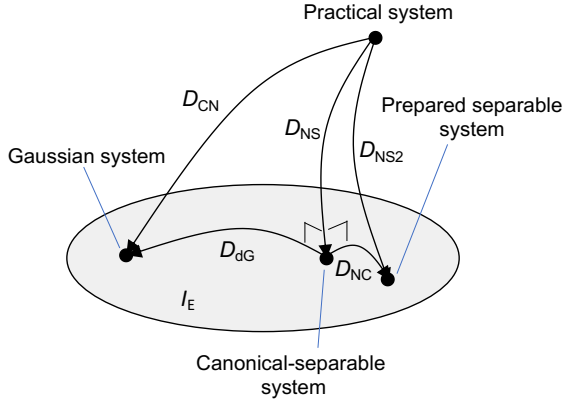}
\caption{ Geometric interpretation of CN on statistical manifold on given configuration $\vec{q}$. $I_{\textrm{E}}$ is $e$-flat subspace consisting of separable probability distributions. }
\label{fig:strategy}
\end{center}
\end{figure}

To solve this problem, we prepare ``alternative'' separable system where CDOS is well-defined, providing further physical insight into $D_{\textrm{NS}}$. Figure~\ref{fig:strategy} shows schematic illustration of CN and its partial contribution, with artifically prepared system.
Since $D_{\textrm{NS}}$ corresponds to dual projection on $e$-flat subspace $I_{\textrm{E}}$, for any prepared separable system (PS) $P_{\textrm{s}}\in I_{\textrm{E}}$, the inequality is satisfied:
\begin{eqnarray}
\label{eq:ieq}
D_{\textrm{NS}} \le D_{\textrm{NS2}}.
\end{eqnarray}
To specify PS applying to the thermodynamic interpretaion, the followings are further required:
(i) CDOS for PS should has the same $\Gamma$ as practical and Gaussian system, and (ii) Difference in PS and practical system should reflect changes in inter-constraint for SDFs, because: Our recent work find that average of $D_{\textrm{NS}}$ over configurations exhibit strong positive correlation with geometry of configurational polyhedra, indicating that inter-constraint for SDFs on configuration space plays central role for $D_{\textrm{NS}}$ character. One of the most natural, accessible PS satisfying the above conditions is the system where its CDOS is constructed by the product of marginal CDOS for practical system, which always results in separable canonical distribution. 
Note that $D_{\textrm{NC}}$ corresponds to difference between $D_{\textrm{NS2}}$ and $D_{\textrm{NS}}$, comes from ``non-commutative (NC)'' character of map $\phi$ and separable-operation $\psi$, i.e., 
\begin{eqnarray}
\left[ \psi, \phi \right] g = \psi\cdot \phi \cdot g - \phi\cdot\psi \cdot g \neq 0,
\end{eqnarray}
which generally holds on except for at perfectly random state of $\left[ \psi,\phi \right] = 0$.

In a similar fashion to the case of $D_{\textrm{CN}}$, we consider system thermodynamic and related functions measured from the PS, described by the hat, e.g., $\hat{M}$ to clarify the role of NS in SDFs, and we focus on the $f=2$ SDF systems. The different notifications from $D_{\textrm{CN}}$ to be cared are (i) we would like to extract average (or summed) information about NS along transition path, to make link to the ground-state configuration geometry, and (ii) effectively employing inequality of Eq.~\eqref{eq:ieq}, where information about $D_{\textrm{NS}}$ (not about $D_{\textrm{NS2}}$) should be addressed. With these considerations, we here introduce special average called ``vicinity difference (VD)'' along the transion $A\to B$ for given function $M$, defined as
\begin{eqnarray}
\label{eq:vd}
\Braket{M}_{A\overline{B}} &=& \sum_{\vec{q}_{B}} R\left( \vec{q}_{B}|\vec{q}_{A} \right) M - \sum_{\vec{q}_{A}} R\left( \vec{q}_{A}|\vec{q}_{B} \right) M \nonumber \\
&=& \Braket{M}_{A} - \Braket{M}_{B}.
\end{eqnarray}
When we apply the VD to bath entropy change measured from PS, we get
\begin{widetext}
\begin{eqnarray}
\label{eq:vdsb}
\Braket{\Delta S_{\textrm{b}}}_{A\overline{B}} = D_{\textrm{NS2}}^{A} + D_{\textrm{NS2}}^{B} - \Braket{\beta \hat{F}_{A}}_{B} - \Braket{\beta \hat{F}_{B}}_{A}  - \left\{ \ln\frac{g\left( \vec{q}_{B} \right)}{g_{\textrm{sp}}\left( \vec{q}_{B} \right) }   + \ln\frac{g\left( \vec{q}_{A} \right)}{g_{\textrm{sp}}\left( \vec{q}_{A} \right) } \right\},
\end{eqnarray}
\end{widetext}
where 
\begin{eqnarray}
D_{\textrm{NS2}}^{J} = \sum_{\vec{q}} R\left( \vec{q} | \vec{q}_{J} \right) \ln\frac{R\left( \vec{q} | \vec{q}_{J} \right)}{ R_{\textrm{sp}}\left( \vec{q} | \vec{q}_{J} \right) }
\end{eqnarray}
corresponds to $D_{\textrm{NS2}}$ at configuration $\vec{q}_{J}$ in Fig.~\ref{fig:strategy}, and $g_{\textrm{sp}}$ denotes CDOS for the PS. 

We can see that the last term of r.h.s. in Eq.~\eqref{eq:vdsb} corresponds to stochastic mutual information between SDFs at each configuration , namely
\begin{eqnarray}
 \ln\frac{g\left( \vec{q}_{B} \right)}{g_{\textrm{sp}}\left( \vec{q}_{B} \right) }   + \ln\frac{g\left( \vec{q}_{A} \right)}{g_{\textrm{sp}}\left( \vec{q}_{A} \right) } = i_{A} + i_{B} = i_{A+B},
\end{eqnarray}
which corresponds to partial contribution to $D_{\textrm{NS}}$ or $D_{\textrm{NS2}}$ at perfectly random state. 
(Hereinafter, we employ superscript or subscript description of $M^{A+B}$ or $M_{A+B}$ as $M_{A}+M_{B}$.)
For the third and fourth term in r.h.s. of Eq.~\eqref{eq:vdsb}, we employ the following relationship to rewrite in terms of the VD:
\begin{eqnarray}
\Braket{\beta\hat{F}_{A}}_{A\overline{B}} &=& \beta \hat{F}_{A} - \Braket{\beta\hat{F}_{A}}_{B} \nonumber \\
\Braket{\beta\hat{F}_{B}}_{A\overline{B}} &=& \Braket{\beta\hat{F}_{B}}_{A} - \beta\hat{F}_{B}.
\end{eqnarray}
Substitution the above relationship into Eq.~\eqref{eq:vdsb}, we obtain
\begin{eqnarray}
\Braket{\Delta S_{\textrm{b}}}_{A\overline{B}} = D_{\textrm{NS2}}^{A+B} - i_{A+B} - \beta\hat{F}_{A+B} - \Braket{\beta\Delta\hat{F}}_{A\overline{B}}.
\end{eqnarray}
Then we further applying the followings:
\begin{eqnarray}
\Delta S_{\textrm{b}} - \Delta d_{0} = -\beta \Delta F,
\end{eqnarray}
we finally obtain the thermodynamic and nonlinear relationships analogous to Eq.~\eqref{eq:bridge1}, namely,
\begin{eqnarray}
\label{eq:bridge2}
D_{\textrm{NS2}}^{A+B} - \left[ i_{A+B} + \Braket{\Delta \hat{d}_{0}}_{A\overline{B}} \right] = \beta \hat{F}_{A+B},
\end{eqnarray}
and thus, for original entanglement in SDFs, we obtain the corresponding upper bound:
\begin{eqnarray}
\label{eq:final}
D_{\textrm{NS1}}^{A+B} - \left[ i_{A+B} + \Braket{\Delta \hat{d}_{0}}_{A\overline{B}} \right] \le \beta \hat{F}_{A+B}.
\end{eqnarray}
The above equations certainly tells us that average entanglement in SDFs measured from sum of stochastic mutual information and stochasic eigen nonlinearity at perfectly random state is bounded by average of free energy at initial and final state. The inequality of Eq.~\eqref{eq:final} comes from the noncommutative character of canonical map $\phi$ and separable operation $\psi$, i.e., $\left[ \psi, \phi \right]\neq 0$, which should be further addressed in our future study. 

\section{Conclusions}
We investigate contribution from nonseparable character (entanglement) in SDFs to nonlinearity in canonical ensemble, in terms of various measure of the nonlinearity on configuration space and on statistical manifold, unified through stochastic thermodynamics. We clarify that through the system transition, average entanglement in SDFs (measured from stochastic information about entanglement and nonlinearity at perfectly random state) is bounded by average free energy at initial and final state, where the bound would be further modified by the knowledge of non-commutative character between canonical map and separable-operation for configurational density of states.

\section{Acknowledgement}
This work was supported by Grant-in-Aids for Scientific Research on Innovative Areas on High Entropy Alloys through the grant number JP18H05453 and  from the MEXT of Japan, and Research Grant from Hitachi Metals$\cdot$Materials Science Foundation.

\end{document}